\RequirePackage{fix-cm}
\documentclass[smallcondensed]{svjour3}     
\smartqed  
\pdfoutput=1
\usepackage{graphicx}
\usepackage{amsmath,amssymb}
\begin{document}

\title{Chained Clauser-Horne-Shimony-Holt inequality for Hardy's ladder test of nonlocality
}

\titlerunning{Chained CHSH inequality for Hardy's ladder test}        

\author{Jos\'{e} Luis Cereceda}


\institute{Jos\'{e} Luis Cereceda \at
              Telef\'{o}nica de Espa\~{n}a, Distrito Telef\'{o}nica, Edificio Este 1, 28050, Madrid, Spain \\             
              \email{jl.cereceda@movistar.es}           
}

\date{Received: date / Accepted: date}

\maketitle

\begin{abstract}
Relativistic causality forbids superluminal signaling between distant observers. By exploiting the non-signaling principle, we
derive the exact relationship between the chained Clauser-Horne-Shimony-Holt sum of correlations $\text{CHSH}_K$ and the
success probability $P_K$ associated with Hardy's ladder test of nonlocality for two qubits and $K+1$ observables per qubit.
Then, by invoking the Tsirelson bound for $\text{CHSH}_K$, the derived relationship allows us to establish an upper limit on
$P_K$. In addition, we draw the connection between $\text{CHSH}_K$ and the chained version of the Clauser-Horne (CH) inequality.
\keywords{Hardy's ladder test of nonlocality \and Non-signaling principle \and Chained CHSH and CH inequalities \and Tsirelson's
bound}
\end{abstract}

\section{Introduction}
\label{sec:1}

In 1992, Lucien Hardy \cite{hardy1} gave a new proof of nonlocality without inequalities for two particles that only requires a total of four
dimensions in Hilbert space. He further showed \cite{hardy2} that this proof works for all pure entangled states of two two-state systems
or qubits except for the maximally entangled state. Hardy's proof \cite{hardy2} (which concerns two observables for each qubit) was
subsequently extended to the case in which there are $K+1$ available dichotomic observables per qubit---$A_0, A_1,\ldots,A_K$ for
qubit $A$ and $B_0, B_1,\ldots,B_K$ for qubit $B$, where $K=1,2,3,\ldots$ \cite{hardy3,boschi}. We will refer to this latter proof as Hardy's
ladder test of nonlocality. In order to have a contradiction between quantum mechanics and local realism in Hardy's ladder scenario, the
observables $A_k$ and $B_{k^{\prime}}$ ($k,k^{\prime}=0,1,\ldots,K$) are required to satisfy the following conditions \cite{hardy3,boschi}
\begin{align}
& P_K = P(A_K =+1, B_K =+1) \neq 0,  \label{hcon1} \\
& P(A_k =+1, B_{k-1} =-1) = 0, \quad\text{for $k=1$ to $K$}, \label{hcon2} \\
& P(A_{k-1} =-1, B_k=+1) = 0, \quad\text{for $k=1$ to $K$}, \label{hcon3} \\
& P(A_0 =+1, B_0 =+1) = 0 , \label{hcon4}
\end{align}
where $P(A_k = i, B_{k^{\prime}}=j)$ is the joint conditional probability of obtaining the result $i$ when measuring $A_k$ on qubit $A$
and obtaining the result $j$ when measuring $B_{k^{\prime}}$ on qubit $B$ ($i,j =\pm 1$). According to a local realistic (LR) theory,
fulfillment of the $2K +1$ conditions in \eqref{hcon2}-\eqref{hcon4} necessarily implies that $P_K =0$. Quantum-mechanically, however,
we can have $P_K \neq 0$ while all the other conditions in \eqref{hcon2}-\eqref{hcon4} are satisfied. The success probability $P_K$ of
Hardy's nonlocality argument is sometimes known as the ``Hardy fraction.''

It should be noted that the conditions \eqref{hcon2}-\eqref{hcon4} could not be satisfied strictly in practical experiments due to the difficulty
of experimentally measuring a null event. Indeed, {\it even\/} with perfect measurement apparatus it is not possible to achieve a true ``zero''
value for the various probabilities because the number of measurements in a real experiment is necessarily finite \cite{mermin1,mermin2}.
Moreover, as pointed out in Ref.~\cite{boschi}, to test experimentally Hardy's conditions \eqref{hcon1}-\eqref{hcon4}, inequalities are necessary
in order to make sure that the errors do not wash out the logical conundrum faced by local realism. One such suitable inequality for Hardy's ladder
test is the chained Clauser-Horne-Shimony-Holt-type (CHSH) inequality \cite{pearle,caves,wehner}
\begin{equation}\label{chsh}
\left| \sum_{k=1}^{K} E(A_{k},B_{k-1}) + \sum_{k=1}^{K} E(A_{k-1},B_{k}) + E(A_K,B_K) - E(A_0,B_0) \right| 
\overset{\text{LR}}{\leq} 2K ,
\end{equation}
which holds for any LR theory, with the correlation function $E(A_k, B_{k^{\prime}})$ defined by $E(A_k, B_{k^{\prime}}) = P^{+}
(A_k, B_{k^{\prime}}) - P^{-}(A_k, B_{k^{\prime}})$, where
\begin{align*}
P^{+}(A_k, B_{k^{\prime}}) & = P(A_k = +1, B_{k^{\prime}}=+1) + P(A_k = -1, B_{k^{\prime}}=-1), \\
P^{-}(A_k, B_{k^{\prime}}) & = P(A_k = +1, B_{k^{\prime}}=-1) + P(A_k = -1, B_{k^{\prime}}=+1),
\end{align*}
and $P^{+}(A_k, B_{k^{\prime}}) + P^{-}(A_k, B_{k^{\prime}}) =1$. Notice that the $2K+2$ pairs of observables $(A_0,B_0)$,
$(A_K, B_K)$, $(A_{k-1},B_k)$, $(A_k,B_{k-1})$, $k=1,2,\ldots,K$, occurring on the left-hand side of inequality \eqref{chsh} are precisely those
in Eqs.~\eqref{hcon1}-\eqref{hcon4}. Evidently, for $K=1$, the inequality \eqref{chsh} reduces to the original CHSH inequality \cite{clauser1}.
It is well known that the maximum quantum violation of the CHSH inequality is given by Tsirelson's bound $2\sqrt{2}$ \cite{tsirelson}. Furthermore,
Wehner \cite{wehner} showed that the corresponding Tsirelson bound for the chained CHSH inequality \eqref{chsh} is given by
\begin{multline}\label{wehner}
\left| \sum_{k=1}^{K} E(A_{k},B_{k-1}) + \sum_{k=1}^{K} E(A_{k-1},B_{k}) + E(A_K,B_K) - E(A_0,B_0) \right| \\
\overset{\text{QM}}{\leq} 2(K+1) \cos\frac{\pi}{2(K+1)}.
\end{multline}

Let us denote the sum of correlations on the left-hand side of either \eqref{chsh} or \eqref{wehner} as $\text{CHSH}_K$, that is,
\begin{equation}\label{para}
\text{CHSH}_K \equiv \sum_{k=1}^{K} E(A_{k},B_{k-1}) + \sum_{k=1}^{K} E(A_{k-1},B_{k}) + E(A_K,B_K) - E(A_0,B_0).
\end{equation}

In this paper (Sec.~\ref{sec:2}), we show that, for the case in which the conditions in Eqs.~\eqref{hcon2}-\eqref{hcon4} are met, the whole
$\text{CHSH}_K$ expression \eqref{para} can be written in terms of the Hardy fraction $P_K$ through the simple relation
\begin{equation}\label{cere}
\text{CHSH}_K = 2K + 4P_K, \quad K=1,2,3,\ldots\, .
\end{equation}
Remarkably, as we will see, relation \eqref{cere} follows as a consequence of the non-signaling (NS) principle alone. Every probabilistic theory
respecting the NS principle (including quantum mechanics) should therefore comply with relation \eqref{cere}.\footnote{
In Sec.~\ref{sec:3}, it will be verified that, in fact, the quantum predictions satisfy relation \eqref{cere}.}
This relation was already obtained elsewhere for the simplest case $K=1$ \cite{cere1}. It has also been derived independently (for $K=1$) by
Xiang in Ref.~\cite{xiang}. Relation \eqref{cere} embodies the precise connection between, on the one hand, Hardy's ladder test of nonlocality
based on Eqs.~\eqref{hcon1}-\eqref{hcon4} and, on the other hand, the test of nonlocality based on the generalized CHSH inequality \eqref{chsh}.

An immediate implication of relation \eqref{cere} is that, when the Hardy conditions \eqref{hcon1}-\eqref{hcon4} are fulfilled, the inequality
\eqref{chsh} is violated by an amount $2K + 4P_K \leq 2K$, or $4P_K \leq 0$. It is important to note that this amount is four times bigger than
that obtained for the chained version of the Clauser-Horne (CH) inequality \cite{boschi,mermin1,mermin2,cere2,clauser2}
\begin{multline*}
P(A_K =+1, B_K =+1) - P(A_0 =+1,B_0=+1) \\
- \sum_{k=1}^{K} \big[P(A_{k}=+1,B_{k-1}=-1) + P(A_{k-1}=-1,B_{k}=+1) \big]  \overset{\text{LR}}{\leq} 0,
\end{multline*}
which is the inequality commonly used in the experimental realizations of Hardy's ladder test of nonlocality (see, for example,
Refs.~\cite{boschi,barbieri,vallone,guo}).

In Sec.~\ref{sec:3}, we will establish (see Eq. \eqref{cere3} below) the relationship between $\text{CHSH}_K$ and the sum of probabilities on
the left-hand side of the above CH-type inequality for the general case in which the probabilities are constrained only by the NS principle. The relation
\eqref{cere} and its generalization \eqref{cere3} are the main results of this paper.

Moreover, combining the relation \eqref{cere} and the Tsirelson bound \eqref{wehner} gives us the following upper limit for $P_K$
\begin{equation}\label{ul}
P_K  \overset{\text{QM}}{\leq}\frac{1}{4}\left[ 2(K+1)\cos\frac{\pi}{2(K+1)} - 2K \right] \equiv L_K .
\end{equation}
Therefore, if quantum mechanics is correct, the Hardy fraction has to be bounded above by the upper limit $L_K$ in Eq.~\eqref{ul}. In particular, for
the first five values of $K$, from \eqref{ul} we obtain
\begin{align*}
\quad P_1 & \leq \tfrac{1}{2}(\sqrt{2} - 1) \approx 0.207106,   &  P_2 & \leq \tfrac{1}{4}(3\sqrt{3} - 4) \approx 0.299038,  \\
\quad P_3 & \leq \sqrt{2 +\sqrt{2}} - \tfrac{3}{2} \approx 0.347759,   &   P_4 & \leq \tfrac{5}{8} \sqrt{10 +2\sqrt{5}} - 2 \approx 0.377641, \\
\quad P_5 & \leq \tfrac{1}{4}\big( 3(\sqrt{2} + \sqrt{6}) -10\big) \approx 0.397777.
\end{align*}

\begin{figure}[ttt]
\begin{center}
\vspace{-.3cm}
\scalebox{0.40}{\includegraphics{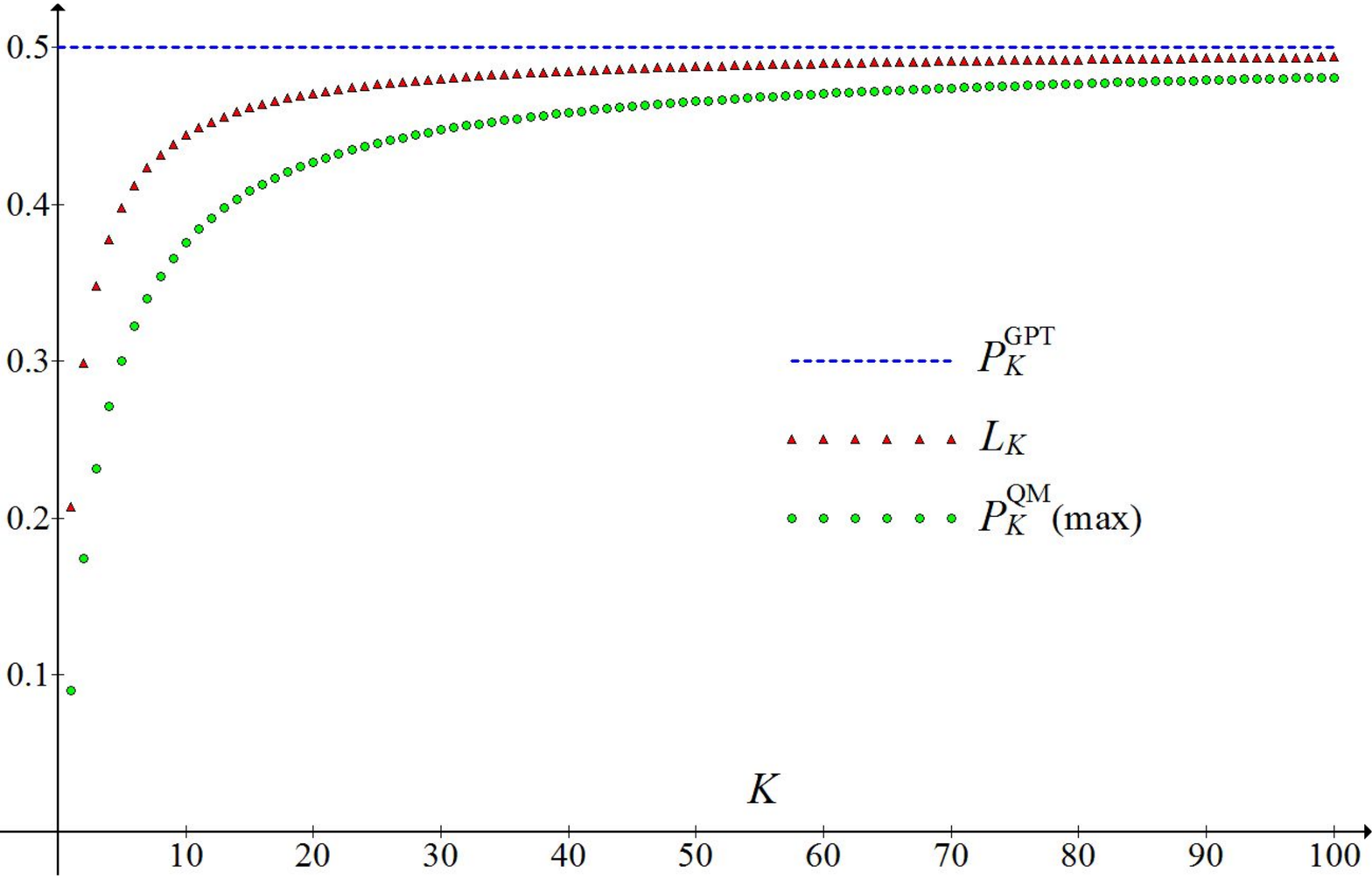}}
\vspace{-.3cm}
\caption{\label{fig:1} Plot of the upper limit $L_K$ (red triangles) and the quantum prediction $P_K^{\text{QM}}(\text{max})$ (green circles) against
$K$ (for $K=1$ to $100$) of the success probability of Hardy's nonlocality argument. The horizontal line $P_K^{\text{GPT}} = 0.5$ represents the maximum
success probability allowed by a generalized probabilistic theory.}
\vspace{-.2cm}
\end{center}
\end{figure}

In Fig.~\ref{fig:1}, we have plotted the upper limit $L_K$ for $K=1$ to $100$. For comparison, we have also plotted the maximum success probability
$P_K^{\text{QM}}(\text{max})$ achieved by quantum mechanics. For a given $K$, this latter probability is obtained by maximizing the function
$P_K^{\text{QM}}(x)$ with respect to $x$ in the interval $0 \leq x \leq 1$, where \cite{hardy3,boschi,cere3}
\begin{equation*}
 P_K^{\text{QM}}(x) = \frac{x^2}{1+x^2}\left( \frac{1-x^{2K}}{1+x^{2K+1}} \right)^2 .
\end{equation*}
The asymptotic value of both $L_K$ and $P_K^{\text{QM}}(\text{max})$ is $L_K = P_K^{\text{QM}}(\text{max}) = 0.5$ in the limit $K \to \infty$.
Quantum-mechanically, the absolute maximum $P_K^{\text{QM}}(\text{max}) = 0.5$ is realized for $K \to \infty$ and a state that is close to maximally
entangled ($ x\to 1$) \cite{hardy3,boschi}. From relation \eqref{cere} it follows that, as $P_K^{\text{QM}}(\text{max}) \to 0.5$, the $\text{CHSH}_K$
expression \eqref{para} approaches the algebraic limit $2K +2$.\footnote{
Incidentally, it is worth pointing out that, in this limit, a direct (``all or nothing'') contradiction between quantum mechanics and local realism emerges in
Hardy's ladder scenario \cite{cere3}.} Furthermore, it is known \cite{cere4} that a generalized probabilistic theory (GPT) adhering to the NS principle allows for a
maximum Hardy's fraction equal to $0.5$ {\it independently\/} of the value of $K$. This is indicated in Fig.~\ref{fig:1} by the horizontal line $P_K^{\text{GPT}}
= 0.5$.\footnote{
Note that, from relation \eqref{cere}, the limit $P_K = 0.5$ can never be surpassed since this would imply that $\text{CHSH}_K > 2K +2$, which is impossible by
the very definition of $\text{CHSH}_K$.}
Indeed, the following extremal NS probability distribution \cite{cere4,jones}
\begin{equation*}
P(A_k \!=\! i, B_{k^{\prime}}\! = \! j) \! = \!
\begin{cases}
\frac{1}{2},  &\!\!\!\!\text{for $i=j$ and $\forall k,k^{\prime}\in\{0,1,\ldots,K \}$ except for $k=k^{\prime}=0$;} \\
0,  &\!\!\!\!\text{for $i = j$ and $k = k^{\prime}=0$;} \\
0,  &\!\!\!\!\text{for $i\neq j$ and $\forall k,k^{\prime}\in\{0,1,\ldots,K \}$ except for $k=k^{\prime}=0$;} \\
\frac{1}{2},  &\!\!\!\!\text{for $i \neq j$ and $k = k^{\prime}=0$,} \\
\end{cases}
\end{equation*}
satisfies all the Hardy conditions \eqref{hcon1}-\eqref{hcon4} with $P_K = 0.5$, as well as the requirements of normalization and non-negativity, and gives the
maximum algebraic bound (namely, $2K +2$) of inequality \eqref{chsh}. For $K=1$, it corresponds to the Popescu-Rohrlich-type correlations \cite{popescu}
leading to the maximum algebraic violation (namely, $4$) of the CHSH inequality while preserving relativistic causality (see Ref.~\cite{fritz} for several variants
of the above extremal distribution for $K=1$).

\section{Chained CHSH Inequality for Hardy's Ladder Test of Nonlocality}
\label{sec:2}

We devote this section to prove relation \eqref{cere}. This is done by employing certain judiciously chosen relationships imposed by the NS principle. For
the Hardy ladder scenario, this principle requires that the marginal probability $P(A_k =i)$ [$P(B_{k^{\prime}}=j)$] of obtaining the result $i$ [$j$] in a
measurement of $A_k$ [$B_{k^{\prime}}$] on qubit $A$ [$B$] is independent of which measurement $B_0, B_1,\ldots,B_K$ [$A_0,A_1,\ldots,A_K$] is
performed on the distant qubit $B$ [$A$]. In terms of joint probabilities, this requirement amounts to the following set of conditions:
\begin{align}
\sum_{j=\pm 1} P(A_k =i, B_0=j) & = \sum_{j=\pm 1} P(A_k =i, B_1=j) \notag  \\
& = \ldots =\sum_{j=\pm 1} P(A_k =i, B_K=j) \quad  \forall k,i,  \label{c1} \\
\sum_{i=\pm 1} P(A_0 =i, B_{k^{\prime}}=j) & = \sum_{i=\pm 1} P(A_1 =i, B_{k^{\prime}}=j) \notag \\
& = \ldots =\sum_{i=\pm 1} P(A_K =i, B_{k^{\prime}}=j) \quad  \forall k^{\prime},j,  \label{c2}
\end{align}
where $k,k^{\prime}=0,1,\ldots,K$ and $i,j = \pm 1$.

To prove relation \eqref{cere}, we first rewrite the $\text{CHSH}_K$ expression \eqref{para} in the equivalent form
\begin{align*}
\text{CHSH}_K = 2K + 2P^{+}(A_K, B_K) & - 2P^{+}(A_0, B_0) \\
& - 2\sum_{k=1}^{K} P^{-}(A_{k},B_{k-1}) - 2\sum_{k=1}^{K} P^{-}(A_{k-1},B_{k}).
\end{align*}
For the case in which the conditions in \eqref{hcon2}-\eqref{hcon4} are fulfilled, the above expression reduces to
\begin{multline*}
\text{CHSH}_K = 2K + 2P_K + 2 P(A_K =-1,B_K =-1) -2P(A_0=-1,B_0 =-1) \\
- 2\sum_{k=1}^{K} P(A_{k}=-1,B_{k-1}=+1) - 2\sum_{k=1}^{K} P(A_{k-1}=+1,B_{k}=-1).
\end{multline*}
Therefore, in order to prove relation \eqref{cere}, it suffices to show that
\begin{multline}\label{cere2}
P(A_K =-1,B_K =-1) = P_K + P(A_0=-1,B_0 =-1) \\
+ \sum_{k=1}^{K} P(A_{k}=-1,B_{k-1}=+1) + \sum_{k=1}^{K} P(A_{k-1}=+1,B_{k}=-1) .
\end{multline}

In what follows we show that relation \eqref{cere2} is indeed fulfilled for $K=1$ and $2$, and then we establish the result generally. In the rest of this section,
we employ the abbreviated notation $P_{k k^{\prime}}^{ij}$ to refer to the joint probability $P(A_k = i,B_{k^{\prime}}=j)$.

\subsection{Case $K=1$}

For $K=1$, the NS conditions in Eqs. \eqref{c1} and \eqref{c2} read as follows:
\begin{equation}
\begin{aligned}\label{K11}
& P_{00}^{++}+ P_{00}^{+-} =  P_{01}^{++}+ P_{01}^{+-}  \\
& P_{10}^{++}+ P_{10}^{+-} =  P_{11}^{++}+ P_{11}^{+-} \\
& P_{00}^{-+}+ P_{00}^{--} =  P_{01}^{-+}+ P_{01}^{--} \\
& P_{10}^{-+}+ P_{10}^{--} =  P_{11}^{-+}+ P_{11}^{--}  \\
& P_{00}^{++}+ P_{00}^{-+} =  P_{10}^{++}+ P_{10}^{-+} \\
& P_{01}^{++}+ P_{01}^{-+} =  P_{11}^{++}+ P_{11}^{-+} \\
& P_{00}^{+-}+ P_{00}^{--} =  P_{10}^{+-}+ P_{10}^{--}  \\
& P_{01}^{+-}+ P_{01}^{--} =  P_{11}^{+-}+ P_{11}^{--} .
\end{aligned}
\end{equation}
Furthermore, Hardy's conditions \eqref{hcon2}-\eqref{hcon4} for $K=1$ mean that
\begin{equation}\label{K12}
P_{00}^{++} = P_{01}^{-+} = P_{10}^{+-} = 0.
\end{equation}
Hence using \eqref{K12} in \eqref{K11}, we readily obtain
\begin{equation}
\begin{aligned}\label{K13}
P_{00}^{+-} =  P_{01}^{++}+ P_{01}^{+-} &  \\
P_{10}^{++} =  P_{11}^{++}+ P_{11}^{+-} & \\
P_{01}^{--} =  P_{00}^{-+}+ P_{00}^{--} & \\
P_{11}^{--}+ P_{11}^{-+} =  P_{10}^{-+}+ P_{10}^{--} &  \\
P_{00}^{-+} =  P_{10}^{++}+ P_{10}^{-+} & \\
P_{01}^{++} =  P_{11}^{++}+ P_{11}^{-+} & \\
P_{10}^{--} =  P_{00}^{+-}+ P_{00}^{--}  & \\
P_{11}^{--}+ P_{11}^{+-} =  P_{01}^{+-}+ P_{01}^{--} & .
\end{aligned}
\end{equation}
Summing all eight relationships in \eqref{K13} and simplifying gives $P_{11}^{--} = P_{11}^{++} + P_{00}^{--} + P_{10}^{-+} +
P_{01}^{+-}$, which is just relation \eqref{cere2} for $K=1$.

\subsection{Case $K=2$}

For $K=2$, the NS conditions in Eqs. \eqref{c1} and \eqref{c2} imply that
\begin{equation}\label{K21}
\begin{aligned}
& P_{00}^{++}+ P_{00}^{+-} = P_{01}^{++}+ P_{01}^{+-} = P_{02}^{++}+ P_{02}^{+-}  \\
& P_{10}^{++}+ P_{10}^{+-} = P_{11}^{++}+ P_{11}^{+-} = P_{12}^{++}+ P_{12}^{+-} \\
& P_{20}^{++}+ P_{20}^{+-} = P_{21}^{++}+ P_{21}^{+-} = P_{22}^{++}+ P_{22}^{+-}  \\
& P_{00}^{-+}+ P_{00}^{--} = P_{01}^{-+}+ P_{01}^{--} = P_{02}^{-+}+ P_{02}^{--}  \\
& P_{10}^{-+}+ P_{10}^{--} = P_{11}^{-+}+ P_{11}^{--} = P_{12}^{-+}+ P_{12}^{--}  \\
& P_{20}^{-+}+ P_{20}^{--} = P_{21}^{-+}+ P_{21}^{--} = P_{22}^{-+}+ P_{22}^{--}  \\
& P_{00}^{++}+ P_{00}^{-+} = P_{10}^{++}+ P_{10}^{-+} = P_{20}^{++}+ P_{20}^{-+}  \\
& P_{01}^{++}+ P_{01}^{-+} = P_{11}^{++}+ P_{11}^{-+} = P_{21}^{++}+ P_{21}^{-+}  \\
& P_{02}^{++}+ P_{02}^{-+} = P_{12}^{++}+ P_{12}^{-+} = P_{22}^{++}+ P_{22}^{-+}  \\
& P_{00}^{+-}+ P_{00}^{--} = P_{10}^{+-}+ P_{10}^{--} = P_{20}^{+-}+ P_{20}^{--}  \\
& P_{01}^{+-}+ P_{01}^{--} = P_{11}^{+-}+ P_{11}^{--} = P_{21}^{+-}+ P_{21}^{--}  \\
& P_{02}^{+-}+ P_{02}^{--} = P_{12}^{+-}+ P_{12}^{--} = P_{22}^{+-}+ P_{22}^{--},
\end{aligned}
\end{equation}
while Hardy's conditions \eqref{hcon2}-\eqref{hcon4} for $K=2$ are
\begin{equation}\label{K22}
P_{00}^{++} = P_{01}^{-+} = P_{10}^{+-} = P_{12}^{-+} = P_{21}^{+-} =0.
\end{equation}
Then, taking into account \eqref{K22}, we pick out the following subset of relationships among those in the set \eqref{K21}:
\begin{equation}
\begin{aligned}\label{K23}
P_{00}^{+-} & = P_{01}^{++}+ P_{01}^{+-}   \\
P_{10}^{++} & =  P_{12}^{++}+ P_{12}^{+-} \\
P_{21}^{++} & = P_{22}^{++}+ P_{22}^{+-}  \\
P_{01}^{--} & = P_{00}^{-+}+ P_{00}^{--}  \\
P_{12}^{--} & = P_{10}^{-+}+ P_{10}^{--}  \\
P_{22}^{--}+ P_{22}^{-+} & = P_{21}^{-+}+ P_{21}^{--} \\
P_{00}^{-+} & = P_{10}^{++}+ P_{10}^{-+}  \\
P_{01}^{++} & = P_{21}^{++}+ P_{21}^{-+}  \\
P_{12}^{++} & = P_{22}^{++}+ P_{22}^{-+}  \\
P_{10}^{--} & = P_{00}^{+-}+ P_{00}^{--}  \\
P_{21}^{--} & = P_{01}^{+-}+ P_{01}^{--}  \\
P_{22}^{--}+ P_{22}^{+-} & = P_{12}^{+-}+ P_{12}^{--} .
\end{aligned}
\end{equation}
Summing all twelve relationships in \eqref{K23} and simplifying, we get $P_{22}^{--} = P_{22}^{++} + P_{00}^{--} + P_{10}^{-+} +
P_{21}^{-+}  + P_{01}^{+-} + P_{12}^{+-}$, which is just relation \eqref{cere2} for $K=2$.

\subsection{The General Case}

The above proofs for $K=1$ and $2$ generalize straightforwardly to an arbitrary number $K+1$ of observables per qubit. To show this,
we write the NS conditions \eqref{c1} and \eqref{c2} in the expanded form
\begin{equation}\label{KK1}
\begin{aligned}
& \! K+1 \left\{
\begin{array}{l}
P_{00}^{++}+ P_{00}^{+-} = P_{01}^{++}+ P_{01}^{+-} = P_{02}^{++}+ P_{02}^{+-} = \ldots =P_{0K}^{++}+ P_{0K}^{+-} \\
P_{10}^{++}+ P_{10}^{+-} = P_{11}^{++}+ P_{11}^{+-} = P_{12}^{++}+ P_{12}^{+-} = \ldots =P_{1K}^{++}+ P_{1K}^{+-} \\
\,\,\,\,\vdots  \\
P_{K0}^{++}+ P_{K0}^{+-} = P_{K1}^{++}+ P_{K1}^{+-} = P_{K2}^{++}+ P_{K2}^{+-} = \ldots =P_{KK}^{++}+ P_{KK}^{+-} \\
\end{array}\right. \\[.2cm]
& \! K+1 \left\{
\begin{array}{l}
P_{00}^{-+}+ P_{00}^{--} = P_{01}^{-+}+ P_{01}^{--} = P_{02}^{-+}+ P_{02}^{--} = \ldots =P_{0K}^{-+}+ P_{0K}^{--} \\
P_{10}^{-+}+ P_{10}^{--} = P_{11}^{-+}+ P_{11}^{--} = P_{12}^{-+}+ P_{12}^{--} = \ldots =P_{1K}^{-+}+ P_{1K}^{--} \\
\,\,\,\,\vdots  \\
P_{K0}^{-+}+ P_{K0}^{--} = P_{K1}^{-+}+ P_{K1}^{--} = P_{K2}^{-+}+ P_{K2}^{--} = \ldots =P_{KK}^{-+}+ P_{KK}^{--} \\
\end{array}\right. \\[.2cm]
& \! K+1 \left\{
\begin{array}{l}
P_{00}^{++}+ P_{00}^{-+} = P_{10}^{++}+ P_{10}^{-+} = P_{20}^{++}+ P_{20}^{-+} = \ldots =P_{K0}^{++}+ P_{K0}^{-+} \\
P_{01}^{++}+ P_{01}^{-+} = P_{11}^{++}+ P_{11}^{-+} = P_{21}^{++}+ P_{21}^{-+} = \ldots =P_{K1}^{++}+ P_{K1}^{-+} \\
\,\,\,\,\vdots  \\
P_{0K}^{++}+ P_{0K}^{-+} = P_{1K}^{++}+ P_{1K}^{-+} = P_{2K}^{++}+ P_{2K}^{-+} = \ldots =P_{KK}^{++}+ P_{KK}^{-+} \\
\end{array}\right. \\[.2cm]
& \! K+1 \left\{
\begin{array}{l}
P_{00}^{+-}+ P_{00}^{--} = P_{10}^{+-}+ P_{10}^{--} = P_{20}^{+-}+ P_{20}^{--} = \ldots =P_{K0}^{+-}+ P_{K0}^{--} \\
P_{01}^{+-}+ P_{01}^{--} = P_{11}^{+-}+ P_{11}^{--} = P_{21}^{+-}+ P_{21}^{--} = \ldots =P_{K1}^{+-}+ P_{K1}^{--} \\
\,\,\,\,\vdots  \\
P_{0K}^{+-}+ P_{0K}^{--} = P_{1K}^{+-}+ P_{1K}^{--} = P_{2K}^{+-}+ P_{2K}^{--} = \ldots =P_{KK}^{+-}+ P_{KK}^{--}
\end{array}\right.
\end{aligned}
\end{equation}
with a total of $4(K+1)$ rows and $K$ equals signs in each row. Furthermore, Hardy's conditions in \eqref{hcon2}-\eqref{hcon4} read as
\begin{equation}\label{KK2}
P_{00}^{++} = P_{01}^{-+} = P_{10}^{+-} = P_{12}^{-+} = P_{21}^{+-} = \ldots = P_{K-1,K}^{-+} = P_{K,K-1}^{+-} = 0.
\end{equation}
Then, using the $2K +1$ conditions \eqref{KK2} in \eqref{KK1}, we can extract the following appropriate set of $4(K+1)$ relationships:
\begin{subequations}
\begin{equation}\label{KK3}
\begin{aligned}
& K+1 \,
\begin{cases}
& \!\!\!\!\!\! P_{00}^{+-} = P_{01}^{++}+ P_{01}^{+-} \\
& \!\!\!\!\!\! P_{10}^{++} = P_{12}^{++}+ P_{12}^{+-}  \\
& \!\!\!\!\!\! \quad\vdots  \\
& \!\!\!\!\!\! P_{K K-1}^{++} = P_{K K}^{++}+ P_{K K}^{+-} 
\end{cases} \\
\vspace{.3cm}
& K+1 \,
\begin{cases}
& \!\!\!\!\!\! P_{01}^{--} = P_{00}^{-+}+ P_{00}^{--} \\
& \!\!\!\!\!\! P_{12}^{--} = P_{10}^{-+}+ P_{10}^{--}  \\
& \!\!\!\!\!\! \quad\vdots   \\
& \!\!\!\!\!\! P_{K-1 K}^{--} = P_{K-1 K-2}^{-+}+ P_{K-1 K-2}^{--}  \\
& \!\!\!\!\!\! P_{K K}^{--}+ P_{K K}^{-+} = P_{K K-1}^{-+}+ P_{K K-1}^{--}
\end{cases}
\end{aligned}
\end{equation}
and
\begin{equation}\label{KK4}
\begin{aligned}
& K+1 \,
\begin{cases}
& \!\!\!\!\!\! P_{00}^{-+} = P_{10}^{++}+ P_{10}^{-+} \\
& \!\!\!\!\!\! P_{01}^{++} = P_{21}^{++}+ P_{21}^{-+}  \\
& \!\!\!\!\!\! \quad\vdots  \\
& \!\!\!\!\!\! P_{K-1 K}^{++} = P_{K K}^{++}+ P_{K K}^{-+} 
\end{cases}  \\
\vspace{.3cm}
& K+1 \,
\begin{cases}
& \!\!\!\!\!\! P_{10}^{--} = P_{00}^{+-}+ P_{00}^{--} \\
& \!\!\!\!\!\! P_{21}^{--} = P_{01}^{+-}+ P_{01}^{--}  \\
& \!\!\!\!\!\! \quad\vdots   \\
& \!\!\!\!\!\! P_{K K-1}^{--} = P_{K-2 K-1}^{+-}+ P_{K-2 K-1}^{--}  \\
& \!\!\!\!\!\! P_{K K}^{--}+ P_{K K}^{+-} = P_{K-1 K}^{+-}+ P_{K-1 K}^{--}
\end{cases}
\end{aligned}
\end{equation}
\end{subequations}
such that the right-hand side (rhs) of each of the $2K+2$ relationships in both subsets \eqref{KK3} and \eqref{KK4} contains exactly one of the
$2K+2$ joint probabilities appearing on the rhs of relation \eqref{cere2}, with each subset exhausting all these $2K+2$ probabilities. In particular,
the probability $P_{K K}^{++}$ [$P_{00}^{--}$] occurs on the rhs of the \mbox{$(K+1)$-th} [\mbox{($K+2)$-th}] relationship in each subset.
On the other hand, the probability $P_{K K}^{--}$ on the left-hand side (lhs) of relation \eqref{cere2} occurs on the lhs of the $(2K+2)$-th relationship
of both subsets  \eqref{KK3} and \eqref{KK4}. Moreover, such subsets exhibit the following additional features: (i) Every single probability
$P_{k k^{\prime}}^{ij}$ on the lhs of the $2K+2$ relationships in \eqref{KK3} [\eqref{KK4}] (leaving aside the distinguished probability
$P_{K K}^{--}$) has the counterpart $P_{k^{\prime}k}^{ji}$ in the rhs of another relationship of the same subset \eqref{KK3} [\eqref{KK4}];
(ii) The corresponding $r$-th relationships in \eqref{KK3} and \eqref{KK4}, $r =1,2,\ldots,2K+2$, can be transformed into each other by swapping
the superscripts $i \leftrightarrow j$ and the subscripts $ k \leftrightarrow k^{\prime}$ of all the probabilities $P_{k k^{\prime}}^{ij}$ entering the
given $r$-th relationships. With all these ingredients at hand, it is not difficult to see that, on summing all $4(K+1)$ relationships in \eqref{KK3} and
\eqref{KK4}, and simplifying, we end up with relation \eqref{cere2}.

\section{Concluding Remarks}
\label{sec:3}

As pointed out in Sec.~\ref{sec:1}, the quantum mechanical predictions should satisfy relation \eqref{cere}, since it is a consequence of the NS principle.
Next we confirm that, indeed, the joint probabilities predicted by quantum mechanics for Hardy's ladder scenario satisfy the equivalent relation \eqref{cere2}.

Consider two qubits $A$ and $B$ in the generic pure entangled state
\begin{equation}\label{state}
|\Psi \rangle = \frac{x}{\sqrt{1+ x^2}} |+\rangle_A |+\rangle_B -  \frac{1}{\sqrt{1+ x^2}} |-\rangle_A |-\rangle_B ,
\end{equation}
where $\{ |+\rangle_A, |-\rangle_A \}$ ($\{ |+\rangle_B, |-\rangle_B \}$) is an arbitrary orthonormal basis in the state space of qubit $A$ ($B$),
and $0 \leq x \leq 1$. Note that $x=0$ ($x=1$) corresponds to the product (maximally entangled) state. For the state \eqref{state} and for an optimal
choice of observables, the quantum prediction (subject to the fulfillment of conditions \eqref{hcon2}-\eqref{hcon4}) for the various joint probabilities in
relation \eqref{cere2} is \cite{hardy3,boschi,cere3}
\begin{align*}
& P^{\text{QM}}(A_0=-1,B_0=-1) = \frac{(1-x)^2}{1+x^2},  \\
& P^{\text{QM}}(A_K=-1,B_K=-1) = \frac{1}{1+x^2}\left( \frac{1-x^{2K+2}}{1+x^{2K+1}} \right)^2 ,  \\
& P^{\text{QM}}(A_K=+1,B_K=+1) = \frac{x^2}{1+x^2}\left( \frac{1-x^{2K}}{1+x^{2K+1}} \right)^2 , \\
\intertext{and, for $k=1,2,\ldots,K$,}
& P^{\text{QM}}(A_k=-1,B_{k-1}=+1) = P^{\text{QM}}(A_{k-1}=+1,B_{k}=-1) \\
& \qquad\qquad\qquad\qquad\qquad\qquad\, = \frac{(1-x^2)^2}{x(1+x^2)} \frac{x^{2k}}{(1+ x^{2k-1})(1+ x^{2k+1})}.
\end{align*}
Substituting these expressions into relation \eqref{cere2} and simplifying it, eventually yields the identity
\begin{equation*}
\sum_{k=1}^{K} \frac{x^{2k}}{(1+ x^{2k-1})(1+ x^{2k+1})} = \frac{x^2 (x^{2K} -1 )}{(1+x)(x^2 -1)(1+x^{2K+1})},
\end{equation*}
which can be easily proved by mathematical induction on $k$. It is worth noting that, by using the auxiliary identity $x^{2K} -1 = (x^2 -1)
\sum_{j=0}^{K-1} x^{2j}$, the above identity can be rewritten as
\begin{equation*}
\sum_{k=1}^{K} \frac{x^{2k}}{(1+ x^{2k-1})(1+ x^{2k+1})} = \frac{1}{(1+x)(1+x^{2K+1})}
\sum_{k=1}^{K} x^{2k},
\end{equation*}
which holds for any real number $x$. Half joking, all in earnest, one could say that previous identity is a nice gift from the NS principle.

On the other hand, it is important to note that, by adding the following two chained CH-type inequalities
\begin{subequations}
\begin{multline}\label{ch1}
P(A_K =+1,B_K=+1) - P(A_0 =+1,B_0=+1) \\
 - \sum_{k=1}^{K} \big[P(A_{k}=+1,B_{k-1}=-1) + P(A_{k-1}=-1,B_{k}=+1) \big]  \overset{\text{LR}}{\leq} 0,
\end{multline}
and
\begin{multline}\label{ch2}
P(A_K =-1,B_K=-1) -  P(A_0 =-1,B_0=-1) \\
 - \sum_{k=1}^{K} \big[P(A_{k}=-1,B_{k-1}=+1) + P(A_{k-1}=+1,B_{k}=-1) \big]  \overset{\text{LR}}{\leq} 0,
\end{multline}
\end{subequations}
we obtain the inequality
\begin{equation*}
P^{+}(A_K,B_K) - P^{+}(A_0,B_0) - \sum_{k=1}^{K} \big[P^{-}(A_{k},B_{k-1}) + P^{-}(A_{k-1},B_{k}) \big]
\overset{\text{LR}}{\leq} 0,
\end{equation*}
which, in turn, can be readily converted into the chained CHSH-type inequality $\text{CHSH}_K  \overset{\text{LR}}{\leq} 2K$, and vice
versa. Notice that the amount of violation of {\it both\/} CH inequalities \eqref{ch1} and \eqref{ch2} when the Hardy conditions
\eqref{hcon1}-\eqref{hcon4} are fulfilled is $P_K \leq 0$. (For the inequality \eqref{ch2}, this follows at once from relation \eqref{cere2}.)

Consider now the sum of probabilities on the lhs of inequality \eqref{ch1}
\begin{multline*}
\text{CH}_K \equiv P_K - P(A_0 =+1,B_0=+1) \\
 - \sum_{k=1}^{K} \big[P(A_{k}=+1,B_{k-1}=-1) + P(A_{k-1}=-1,B_{k}=+1) \big] ,
\end{multline*}
where we assume that the various probabilities $P(A_k = i,B_{k^{\prime}}=j)$ satisfy the NS conditions \eqref{c1} and \eqref{c2} (apart
from the usual non-negativity and normalization constraints), but are otherwise arbitrary. Then, by applying a procedure similar to that used in
Sec.~\ref{sec:2} to prove relation \eqref{cere}, it can be shown that
\begin{equation}\label{cere3}
\text{CHSH}_K = 2K + 4 \text{CH}_K,  \quad K=1,2,3,\ldots\, .
\end{equation}
From the Tsirelson bound \eqref{wehner}, we therefore deduce that
\begin{equation}\label{ul2}
\text{CH}_K  \overset{\text{QM}}{\leq} L_K,
\end{equation}
where $L_K$ is the upper limit in Eq. \eqref{ul}. In particular, for $K=1$, we retrieve the well-known quantum mechanical bound $\text{CH}_1 
\overset{\text{QM}}{\leq} \frac{1}{2}(\sqrt{2} -1)$ \cite{lima}. Evidently, when the Hardy conditions \eqref{hcon2}-\eqref{hcon4} are satisfied,
the relations \eqref{cere3} and \eqref{ul2} reduce to relations \eqref{cere} and \eqref{ul}, respectively.

Lastly, it is to be mentioned that Ahanj {\it et al}. \cite{ahanj} (see also Ref.~\cite{xiang}) derived an upper bound on the Hardy fraction, for $K=1$,
by applying a sufficient condition for violating the principle of information causality \cite{pawlowski} (IC).\footnote{
The IC principle states that communication of $m$ classical bits causes information gain of at most $m$ bits. The NS principle is just IC for $m=0$.}
Under this condition, they found an upper bound given by $P_1 \leq \frac{1}{2}(\sqrt{2} - 1)$. Note that this bound is the same as the resulting
upper limit in Eq.~\eqref{ul} for $K=1$. This coincidence, however, is not accidental. Indeed, the link between the two approaches becomes clear upon
considering the following two facts: (i) All NS correlations which violate the Tsirelson bound $2\sqrt{2}$ also violate IC \cite{pawlowski,allcock}; (ii) By
relation \eqref{cere}, we have that $\text{CHSH}_1 > 2\sqrt{2}$ whenever $P_1 > \frac{1}{2}(\sqrt{2} - 1)$ \cite{xiang}. We thus conclude that the
set of NS correlations fulfilling all the Hardy conditions \eqref{hcon1}-\eqref{hcon4} with $\frac{1}{2}(\sqrt{2} - 1) < P_1 \leq  0.5$, violate IC.

%
%



\begin{thebibliography}{99}

\bibitem{hardy1} Hardy, L.: Quantum mechanics, local realistic theories, and Lorentz-invariant realistic theories. Phys. Rev. Lett. {\bf 68},
2981--2984 (1992)

\bibitem{hardy2} Hardy, L.: Nonlocality for two particles without inequalities for almost all entangled states. Phys. Rev. Lett. {\bf 71},
1665--1668 (1993)

\bibitem{hardy3} Hardy, L.: A bigger contradiction between quantum theory and locality for two particles without inequalities. In Ferrero M.,
Van der Merwe A. (eds.) New Developments on Fundamental Problems in Quantum Physics, pp.~163--170. Kluwer Academic Publishers,
Dordrecht, The Netherlands (1997)

\bibitem{boschi} Boschi, D., Branca, S., De Martini, F., Hardy, L.: Ladder proof of nonlocality without inequalities: Theoretical
and experimental results. Phys. Rev. Lett. {\bf 79}, 2755--2758 (1997)

\bibitem{mermin1} Mermin, N.D.: Quantum mysteries refined. Am. J. Phys. {\bf 62}, 880--887 (1994)

\bibitem{mermin2} Mermin, N.D.: The best version of Bell's theorem, in Greenberger D.M., Zeilinger A. (eds.) Fundamental Problems in
Quantum Theory: A Conference held in Honor of Professor John A. Wheeler, Ann. N. Y. Acad. Sci. {\bf 755}, pp.~616--623 (1995)

\bibitem{pearle} Pearle, P.M.: Hidden-variable example based upon data rejection. Phys. Rev. D, {\bf 2} 1418--1425 (1970)

\bibitem{caves} Braunstein, S.L., Caves, C.M.: Wringing out better Bell inequalities. Ann. Phys. (N.Y.) {\bf 202}, 22--56 (1990)

\bibitem{wehner} Wehner, S.: Tsirelson bounds for generalized Clauser-Horne-Shimony-Holt inequalities. Phys. Rev. A {\bf 73}, 022110
(2006)

\bibitem{clauser1} Clauser, J.F., Horne, M.A., Shimony, A., Holt, R.A.: Proposed experiment to test local hidden-variable theories. Phys. Rev. Lett.
{\bf 23}, 880--884 (1969)

\bibitem{tsirelson} Cirel'son [Tsirelson], B.S.: Quantum generalizations of Bell's inequality. Lett. Math. Phys. {\bf 4}, 93--100 (1980)

\bibitem{cere1} Cereceda, J.L.: Quantum mechanical probabilities and general probabilistic constraints for Einstein-Podolsky-Rosen-Bohm
experiments. Found. Phys. Lett. {\bf 13}, 427--442 (2000)

\bibitem{xiang} Xiang, Y.: The relation between Hardy's non-locality and violation of Bell inequality. Chin. Phys. B {\bf 20}, 060301 (2011)

\bibitem{cere2} Cereceda, J.L.: Identification of all Hardy-type correlations for two photons or particles with spin 1/2. Found. Phys. Lett.
{\bf 14}, 401--424 (2001)

\bibitem{clauser2} Clauser, J.F., Horne, M.A.: Experimental consequences of objective local theories. Phys. Rev. D {\bf 10}, 526--535 (1974)

\bibitem{barbieri} Barbieri, M., Cinelli, C., De Martini, F., Mataloni, P.: Test of quantum nonlocality by full collection of polarization entangled
photon pairs. Eur. Phys. J. D {\bf 32}, 261--267 (2005)

\bibitem{vallone} Vallone, G., Gianani I., Inostroza, E.B., Saavedra, C., Lima, G., Cabello, A., Mataloni, P.: Testing Hardy's nonlocality proof
with genuine energy-time entanglement. Phys. Rev. A {\bf 83}, 042105 (2011)

\bibitem{guo} Guo, W.J., Fan, D.H., Wei, L.F.: Experimentally testing Bell's theorem based on Hardy's nonlocal ladder proofs. Sci. China Phys.
Mec. {\bf 58}, 1--5 (2015)

\bibitem{cere3} Cereceda, J.L.: Ladder proof of nonlocality for two spin-half particles revisited. J. Phys. A: Math. Gen. {\bf 35}, 9105--9111
(2002)

\bibitem{cere4} Cereceda, J.L.: Causal communication constraint for two qubits in Hardy's ladder proof of non-locality. Adv. Stud. Theor. Phys.
{\bf 9}, 433--448 (2015)

\bibitem{jones} Jones, N.S., Masanes, Ll.: Interconversion of nonlocal correlations. Phys. Rev. A {\bf 72}, 052312 (2005)

\bibitem{popescu} Popescu, S., Rohrlich, D.: Quantum nonlocality as an axiom. Found. Phys. {\bf 24}, 379--385 (1994)

\bibitem{fritz} Fritz, T.: Quantum analogues of Hardy's nonlocality paradox. Found. Phys. {\bf 41}, 1493--1501 (2011)

\bibitem{lima} Lima, G., Inostroza, E.B., Vianna, R.O., Larsson, J.A., Saavedra, C.: Optimal measurement bases for Bell tests based on the
Clauser-Horne inequality. Phys. Rev. A {\bf 85}, 012105 (2012)

\bibitem{ahanj} Ahanj, A., Kunkri, S., Rai, A., Rahaman, R., Joag, P.S.: Bound on Hardy's nonlocality from the principle of information
causality. Phys. Rev. A {\bf 81}, 032103 (2010)

\bibitem{pawlowski} Paw\l{}owski, M., Paterek, T., Kaszlikowski, D., Scarani, V., Winter, A., \.{Z}ukowski, M.: Information causality as a physical
principle, Nature (London) {\bf 461}, 1101--1104 (2009)

\bibitem{allcock} Allcock, J., Brunner, N., Paw\l{}owski, M., Scarani, V.: Recovering part of the quantum boundary from information causality.
Phys. Rev. A {\bf 80}, 040103 (2009)


\end{thebibliography}


\end{document}